\documentclass[useAMS,usenatbib,usegraphicx]{mn2e}
\usepackage{txfonts}
\title[Orbital parameters of BeXRBs in the SMC]{On the orbital parameters of Be/X-ray binaries in the Small Magellanic Cloud}
\author[L. J. Townsend et al.]{L. J. Townsend$^{1}$\thanks{E-mail: ljt203@soton.ac.uk (LJT)}, M. J. Coe$^{1}$, R. H. D. Corbet$^{2}$, A. B. Hill$^{1}$\\
$^{1}$School of Physics and Astronomy, University of Southampton, Highfield, Southampton, SO17 1BJ, United Kingdom\\
$^{2}$University of Maryland Baltimore County, X-ray Astrophysics Laboratory, Mail Code 662; NASA Goddard Space Flight Center, Greenbelt, MD 20771, USA\\}

\begin{document}

\date{Accepted 2011 May 29. Received 2011 May 27; in original form 2011 April 05}

\pagerange{\pageref{firstpage}--\pageref{lastpage}} \pubyear{2011}

\maketitle

\label{firstpage}

\begin{abstract}

\noindent The orbital motion of a neutron star about its optical companion presents a window through which to study the orbital parameters of that binary system. This has been used extensively in the Milky Way to calculate these parameters for several high-mass X-ray binaries. Using several years of \textit{RXTE} PCA data, we derive the orbital parameters of four Be/X-ray binary systems in the SMC, increasing the number of systems with orbital solutions by a factor of three. We find one new orbital period, confirm a second and discuss the parameters with comparison to the Galactic systems. Despite the low metallicity in the SMC, these binary systems sit amongst the Galactic distribution of orbital periods and eccentricities, suggesting that metallicity may not play an important role in the evolution of high-mass X-ray binary systems. A plot of orbital period against eccentricity shows that the supergiant, Be and low eccentricity OB transient systems occupy separate regions of the parameter space; akin to the separated regions on the Corbet diagram. Using a Spearman's rank correlation test, we also find a possible correlation between the two parameters. The mass functions, inclinations and orbital semimajor axes are derived for the SMC systems based on the binary parameters and the spectral classification of the optical counterpart. As a by-product of our work, we present a catalogue of the orbital parameters for every high-mass X-ray binary in the Galaxy and Magellanic Clouds for which they are known.

\end{abstract}


\section{Introduction}

X-ray outbursts in transient high-mass X-ray binaries (HMXB) occur when a compact object accretes intermittently from material being expelled by a massive companion. This intermittent accretion generally occurs because of an eccentric orbit, though transient outbursts from near-circular orbit systems have been seen (e.g. \citealt{pfahl2002}). Classically, such transient outbursts are thought to fall into one of two types. Type I outbursts occur near periastron passage of the neutron star (NS), where there is more accretable material. Studying these outbursts allows orbital periods to be derived through timing analysis of their light curves. Type II outbursts are typically an order of magnitude brighter than Type I outbursts and last much longer. The cause of this being an enlargement of the circumstellar material or radially driven wind around the companion star (e.g. \citealt{neg00}), allowing accretion to occur at any orbital phase and at a higher rate. Both of these transient outbursts are thought to originate from the Be/X-ray binary (BeXRB) class of HMXB in which the optical star is an early type Be or late Oe main sequence star and the compact object is a NS (there are currently no confirmed Be/black-hole binaries; \citealt{bel09}). The NS is usually in a wide, eccentric orbit allowing for long periods of little X-ray activity. A third type of transient X-ray outburst has recently been detected and is now accepted to originate from a new class of HMXB known as supergiant fast X-ray transients (SFXT). These sources are characterised by rapid X-ray outbursts on the order of minutes to hours \citep{sgu05}, large dynamic ranges of order 10$^{4}$--10$^{5}$ and are associated with OB supergiant companion stars \citep{chat08}. The other major sub-class of HMXB are the supergiant X-ray binaries (SGXB) in which a NS or black hole orbit a giant or supergiant star. Here the compact object is in a tight, near circular orbit, allowing for continuous accretion from the stellar wind or via Roche-Lobe overflow. The final sub-class making up the known HMXB family is the small, but growing class of $\gamma$-ray binaries, in which very high energy radiation is thought to be produced via interaction of the pulsar wind with the stellar atmosphere of the secondary star (e.g. \citealt{hill10}).

\newpage

The Small Magellanic Cloud (SMC) holds a disproportionately large number of HMXBs when compared to the Milky Way. Based on their relative masses, there are a factor of 50 more than would be expected. Population synthesis models by \citet{dray06} predict the low metallicity environment in the SMC could increase the number of HMXBs by a factor of 3. However, this is still not sufficient to explain the SMC population on its own. A recent episode of star formation, possibly due to an increase in the tidal force exerted on the SMC by a close approach with the Large Magellanic Cloud (LMC; \citealt{gard96}), is the current favoured scenario to explain the number of binary systems seen. The metallicity difference is of particular relevance to our investigation as it has substantial influence in the evolution of massive stars. The line-driven stellar winds in massive stars are weaker in low metallicity environments (e.g. \citealt{kud87}), resulting in lower mass and angular momentum losses from the binary systems in which they reside. This could result in compact object masses different to those seen in the Galaxy and could even affect the evolutionary path of the binary system. The latter may manifest itself as differences in observable parameters of the population, such as spectral distribution, the distribution of orbital sizes and eccentricities or the distribution of mass functions. \citet{mcbride08} show that, despite the low metallicity in the SMC, the distribution of spectral types of the optical counterparts to Be/X-ray binaries (BeXRB) is consistent with that of Galactic systems.

\begin{table}
  \caption{Outburst date (T$_{0}$), duration and peak X-ray luminosity of the four systems investigated in this study.}
  \label{tab:outbursts}
  \centering
\begin{tabular}{|c|c|c|c|}
  \hline
  Source & T$_{0}$ (MJD) & Duration (days) & Peak L$_{x}$ ($10^{37} \mathrm{ergs\,s^{-1}}$) \\
  \hline
  \hline
  SXP2.37 & 51573 & 84 & 21.0 \\
  SXP6.85 & 55435 & 59 & 3.3 \\
  SXP8.80 & 55080 & 39 & 7.3 \\
  SXP74.7 & 55232 & 42 & 3.5 \\
  \hline
\end{tabular}\\
\end{table}

Our aim is to use several years of data taken with the \textit{Rossi X-ray Timing Explorer (RXTE)} observatory to uncover the first sizable sample of orbital parameters of HMXBs in the SMC. We do this by fitting a radial velocity model to NS spin period variations that are seen during several Type II outbursts. If the variability is due to the motion of the NS around the optical counterpart, we can obtain a best fit solution of the orbital parameters. The model and data are presented in section 2. The nature of this method means a good fit can only be achieved if the outburst lasts for at least one orbit and the observations are made regularly throughout that orbit. In section 3 we show the fits to 4 systems that were possible with our data sample (see \citealt{gal08} for details of our monitoring programme). These results now make 6 systems in the SMC for which we have good orbital solutions (excluding the only known SGXB SMC X-1). In section 4 we discuss these systems in the context of the Galactic population before summarising our results in section 5.

\section{Data \& Orbital Model}

\textit{RXTE} monitoring of the SMC \citep{gal08} has resulted in 10 year X-ray light curves and spin period histories for most of the known pulsars in the SMC. During this time, we have detected 12 systems that have shown evidence for one or more Type II outbursts. Of these systems only a handful were of sufficient length and had frequent enough observations within the outburst to achieve a good model fit. Table \ref{tab:outbursts} gives a summary of the start time, duration and peak X-ray luminosity in the 3--10\,keV band, of the studied outburst for each system. Each observation was made using the \textit{RXTE} PCA in GoodXenon mode. Data were extracted in the 3--10\,keV energy range and binned at 0.01s. The resulting light curves were background subtracted and barycentre corrected before finally correcting the count rate for number of active PCU's. The final light curves were passed through a Lomb-Scargle periodogram to search for pulsations. The error associated with any period found is calculated based on the formula for the standard deviation of the frequency given in \citet{horne86}:

\begin{eqnarray}
 \delta \omega & = & \frac{3\pi \sigma_{N}}{2 N^{1/2} D\,A}
  \label{equ:error}
\end{eqnarray}

\noindent where $\sigma_{N}$\,$^{2}$ is the variance of the light curve, N is the number of data points, \textit{D} is the length of the data and \textit{A} is the amplitude of the signal given by:

\begin{eqnarray}
 A & = & 2\sqrt{\frac{z_{0}{\sigma_{N}}^{2}}{N}}
  \label{equ:sigamplitude}
\end{eqnarray}

\noindent where $z_{0}$ is the Lomb-Scargle power.

\begin{figure}
 \includegraphics[width=64mm,angle=90]{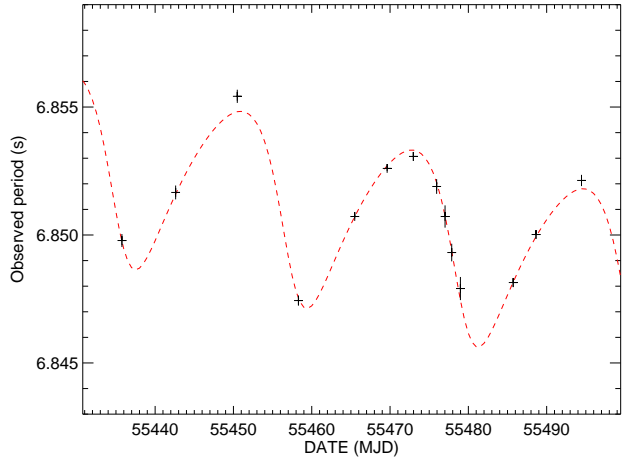}
  \caption{The spin period of SXP6.85 as measured by \textit{RXTE} during 2 months of pointed observations. Both the orbital motion of the NS and the accretion driven spin-up are visible. Overplotted is the model fit to the data as shown in Table \ref{tab:orbit2}.\label{fig:orbit2}}
\end{figure}

In the simplest example of the orbital plane being perpendicular to our line-of-sight (inclination, $i = 0$), any change seen in the spin period of the NS should be due to torque produced by accretion onto the NS surface. The most common result of this is for the spin period to decrease over the length of the outburst. This is known as spin-up. However, in some cases it is seen that the spin period \textit{increases} during phases of accretion. This is known as spin-down. Spin-down also occurs during phases in which there is no accretion due to torques exerted by the magnetic field of the NS. In the more generic case of the orbital plane being at some angle to our line-of-sight, we see more complicated spin period variations. Convolved with the standard spin-up of the NS, there will be a shifting of the X-ray pulse arrival time caused by the orbital motion of the NS around its counterpart. It is because of this that our orbit fitting code simultaneously fits a simple spin-up model and an orbital radial velocity model

\begin{eqnarray}
P_{obs} & = & \left(1 + \frac{v(t)}{c} \right) P(t)
\end{eqnarray}

\noindent where $P_{obs}$ is the detected spin period of the pulsar, $c$ is the speed of light and $v(t)$ is the binary radial velocity. $v(t)$ is calculated using the IDL routine {\sc binradvel.pro} from the {\sc aitlib}\footnote{Developed at the Institut f\"{u}r Astronomie und Astrophysik, Abteilung Astronomie, of the University of T\"{u}bingen and at the University of Warwick, UK; see http://astro.uni-tuebingen.de/software/idl/} which uses the procedure of \citet{2001icbs.book.....H} and employs the method of \citet{1987CeMec..40..329M} to solve Kepler's equation. The spin of the pulsar is given by $P(t)$ which incorporates the spin-up component and is defined below:

\begin{eqnarray}
 P(t) & = & P(t_{0}) + \dot{P}(t - t_{0}) - \frac{\ddot{P}(t - t_{0})^{2}}{2}
  \label{equ:spin}
\end{eqnarray}

\noindent where $\dot{P}$ and $\ddot{P}$ are the spin-up and change in spin-up of the NS. We perform a least-squares fit on the data and return the best fitting parameters. The full spin-up component shown in equation \ref{equ:spin} was fit to every outburst. However, no data set allowed for sufficient statistics in the fit to justify a $\ddot{P}$ component and hence it was removed.

The results of our fits are presented in the next section. In order to obtain a meaningful fit, the period measurements used in the fitting procedure were restricted to detections greater than a particular significance. This threshold was not kept constant for every source because the luminosity and duration of each outburst was different, meaning fits improved or worsened by including data above different thresholds. As such, a significance threshold was chosen to minimise the (1$\sigma$) errors in the binary parameters for each fit. The chosen value is stated in the relevant section.

\begin{table}
  \caption{The orbital parameters for SXP6.85 from the analysis of 3--10 keV \textit{RXTE} PCA data.}
  \label{tab:orbit2}
  \centering
\begin{tabular}{|l|c|l|}
  \hline
  Parameter &  & Orbital Solution \\
  \hline
  \hline
  Orbital period & $\textit{P}_{orbital}$ (d) & $21.9\pm0.1$ \\
  Projected semimajor axis & $\textit{a}_{x}$sin{\it i} (light-s) & $151\pm6$ \\
  Longitude of periastron & $\omega$ ($^{o}$) & $125\pm6$ \\
  Eccentricity & $\textit{e}$ & $0.26\pm0.03$ \\
  Orbital epoch & $\tau_{periastron}$ (MJD) & $55479.1\pm0.4$ \\
  Spin period & $\textit{P}$ (s) & $6.8508\pm0.0001$ \\
  First derivative of $\textit{P}$ & $\dot{P}$ ($10^{-10}\mathrm{ss}^{-1}$) & $-8.0\pm0.5$ \\
  Goodness of fit & $\chi^{2}_{\nu}$ & 1.79 \\
  \hline
\end{tabular}
\end{table}

\section{Results}

\subsection{SXP6.85 = XTE J0103$-$728}

SXP6.85 was first detected in 2003 by \textit{RXTE} as a 6.848\,s pulsed X-ray source \citep{cor03}. It was later detected in a 2006 \textit{XMM-Newton} observation at the position R.A. = $01^{h}02^{m}53\fs1$, dec. = $-72^{\circ}44^{'}33\farcs0$ (J2000.0). This detection led to the identification of a V=14.6 optical counterpart \citep{hab08}, allowing it to be classified as an HMXB. Follow-up work by \citet{mcbride08} classified the counterpart as an O9.5\,V--B0\,V emission line star. In subsequent years it has been detected on 5 distinct occasions, coinciding with times when the counterpart was optically bright \citep{town10}. \citet{kem08} find that the optical flux varies by $\sim$\,0.5 magnitudes with a period of $620 \pm 18$\,d. They associate this with the growth and decay of the circumstellar disk. Those authors also show that the source gets redder as it gets brighter, suggestive of a low inclination system, and propose a low eccentricity based on comparison to other systems.

Until now, the orbital period of this system was not known for certain. Analysis of optical light curves by \citet{kem08} and \citet{schmit07} showed hints of periodicities at $114.1 \pm 0.6$\,d and $24.8 \pm 0.1$\,d, but neither could be confirmed as the orbital period of the system. The 25\,d period was closer to the expected orbital period based on the Corbet diagram \citep{cor86}, but the 114\,d period was reinforced by the detection of a 112\,d period in the X-ray light curve \citep{gal08}. To try and resolve this issue, the two longest Type II X-ray outbursts were fit with our orbital model. The outburst occuring around MJD 54800 turned out to be too sparsely covered, meaning we were unable to get an acceptable fit. However, the outburst beginning on MJD 55435 (Table \ref{tab:outbursts}) was sampled very well thanks to dedicated pointings at the position of the source in addition to our regular monitoring. The period evolution is shown in Fig. \ref{fig:orbit2} with the best model fit overplotted. The data used in the fit were cut at the 99.99\,$\%$ (4\,$\sigma$) significance level. The very clear changes in the spin period allowed the radial velocity of the NS to be found quite simply. Unfortunately, we were unable to fit both outbursts simultaneously as the amount of spin-down happening between the outbursts is unknown. The solution is presented in Table \ref{tab:orbit2} and is the best fit we have in our sample. We propose $21.9\pm0.1$\,d to be the true orbital period of this system, placing it nicely in the BeXRB region of the Corbet diagram. The other orbital parameters and the nature of the 25\,d and 114\,d periods are discussed in section 4.

\subsection{SXP2.37 = SMC X$-$2}

\begin{figure}
 \includegraphics[width=64mm,angle=90]{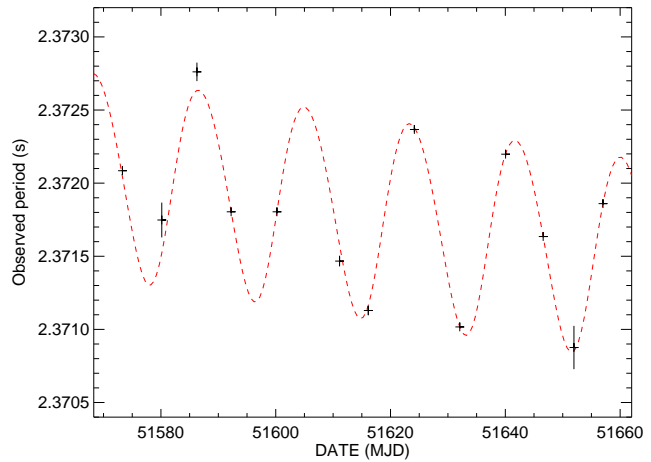}
  \caption{The spin period of SXP2.37 as measured by \textit{RXTE}. The orbital modulation is not as obvious as in the case of SXP6.85, but the period deviations around the general spin-up trend suggests we are seeing orbital modulations. Overplotted is the model fit to the data as shown in Table \ref{tab:orbit4}.\label{fig:orbit4}}
\end{figure}

SXP2.37 was discovered by SAS 3 observations in 1977 \citep{li77} as a highly variable X-ray source. \citet{cor01} discovered 2.372\,s pulsations from the source during the Type II outburst from January 2000 to May 2000. The proposed optical counterpart to this X-ray source was shown to be a close north-south double \citep{mmt79}. As such, it has been difficult to tell which of the two stars is the true optical counterpart due to large X-ray error circles and the lack of any significant periodicity in OGLE II and the first 5 years of OGLE III data \citep{scu06}. It was not until a further 2 years of OGLE III data became available that the correct counterpart revealed itself. \citet{schurch10} show there is a clear $18.62 \pm 0.02$\,d periodicity in the final 2 years of the light curve of the northern star, making it very probable that this star is the true counterpart and that the periodicity is the orbital period of the binary system. \citet{scu09} suggest that the 18.6\,d period, along with the apparent 9 and 6\,d harmonics are caused by 2 nonradial pulsations seen around 0.86\,s and 0.9\,s. \citet{schurch10} argue that because the 18.6\,d modulation is visible in the light curve and the folded light curve at this period is typical of other binary systems, the modulation must be orbital in nature. \citet{mcbride08} classify the counterpart as an O9.5\,III--V emission line star.

\begin{table}
  \caption{The orbital parameters for SXP2.37 from the analysis of 3--10 keV \textit{RXTE} PCA data.}
  \label{tab:orbit4}
  \centering
\begin{tabular}{|l|c|l|}
  \hline
  Parameter &  & Orbital Solution \\
  \hline
  \hline
  Orbital period & $\textit{P}_{orbital}$ (d) & $18.38\pm0.02$ \\
  Projected semimajor axis & $\textit{a}_{x}$sin{\it i} (light-s) & $73.7\pm0.9$ \\
  Longitude of periastron & $\omega$ ($^{o}$) & $226\pm8$ \\
  Eccentricity & $\textit{e}$ & $0.07\pm0.02$ \\
  Orbital epoch & $\tau_{periastron}$ (MJD) & $51616.8\pm0.4$ \\
  Spin period & $\textit{P}$ (s) & $2.37194\pm0.00001$ \\
  First derivative of $\textit{P}$ & $\dot{P}$ ($10^{-10}\mathrm{ss}^{-1}$) & $-0.720\pm0.015$ \\
  Goodness of fit & $\chi^{2}_{\nu}$ & 3.58 \\
  \hline
\end{tabular}
\end{table}

The outburst that led to the discovery of pulsations from SXP2.37 \citep{cor01} is the only time \textit{RXTE} has detected the source with any high significance or for any great period of time. Thus, we tried fitting the period measurements from the outburst with our model. Fits were tried both with the orbital period free to vary and fixed at the value reported in \citet{schurch10}. Both fits yielded the same parameters to within their errors and very similar goodness of fits, verifying that the 18.6\,d periodicity is the orbital period of the system. Fig. \ref{fig:orbit4} shows the pulse period evolution of the source. The data are cut at the 99.9999$\%$ (5\,$\sigma$) significance level. Overplotted is the best model fit to the data in which the orbital period was free to vary. We refine the orbital period measurement to $18.38\pm0.02$\,d. The other parameters are shown in Table \ref{tab:orbit4} and are discussed later.

\begin{figure}
 \includegraphics[width=64mm,angle=90]{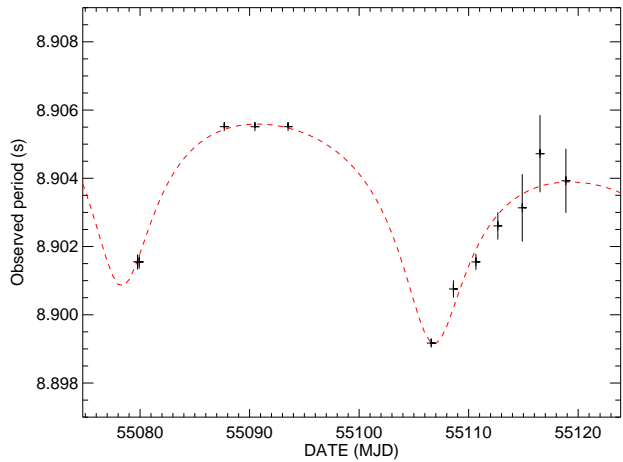}
  \caption{The spin period of SXP8.80 as measured by \textit{RXTE}. The orbital modulation is apparent, though the data are sparse over some of the orbit. The orbital period had to be frozen at 28.51\,d to produce an acceptable fit. This fit is overplotted and is shown in Table \ref{tab:orbit5}.\label{fig:orbit5}}
\end{figure}

\subsection{SXP8.80 = RX J0051.8$-$7231 = 2E 0050.1$-$7247}

SXP8.80 was discovered by \textit{Einstein} IPC observations in 1980 \citep{bruh87} as a new X-ray source in the SMC wing. The original error circle from these observations led to a tentative association with the optical star AzV 111. \citet{wang92} showed that this source was highly variable, had a moderate X-ray flux and a hard spectrum, suggesting a BeXRB classification. \textit{ROSAT} PSPC observations in May 1993 discovered a periodicity of 8.88163 $\pm$ 0.00001\,s in the direction of SXP8.80 \citep{israel95}. Those authors also showed the transient nature of the source which, along with the spin period, confirmed this as a new BeXRB in the SMC. A refined X-ray position published by \citet{kahabka96} cast doubt over the true optical counterpart, with more than one star looking possible. It was not until \citet{hab00} improved the \textit{ROSAT} error circle that the star MA93 506 was identified as the most probable counterpart. The orbital period of SXP8.80 was found to be 28.0 $\pm$ 0.3\,d \citep{cor04} during \textit{RXTE} monitoring of the SMC. Subsequent analysis of MACHO and OGLE data of MA93 506 revealed periodicities of 185\,d and 33\,d respectively (\citealt{coe05}; \citealt{schmit06}), casting doubt over the orbital period and the optical counterpart determination. \citet{mcbride08} classify MA93 506 as a O9.5--B0\,IV--V emission line star. Only recently has this star been verified as the counterpart with the detection of a 28.51 $\pm$ 0.01\,d periodicity in the combined OGLE II \& III light curve \citep{andry11}, in agreement with the X-ray derived period.

SXP8.80 had been X-ray quiet since a long period of Type I outbursts that began around MJD 52700. The Type II outburst to which we have fit our model began on MJD 55080 and is one of the most luminous outbursts seen in the SMC (see Table \ref{tab:outbursts}); 6 or more harmonics of the NS spin period were often seen in the power spectra. Unfortunately, the coverage during this outburst was sparse and so there are large gaps throughout the outburst. The data are shown in Fig. \ref{fig:orbit5} and are cut at the 90\,$\%$ significance level to try and include as much data as possible. This means that some of the data toward the end of the outburst have large associated errors. To aid fitting further, the orbital period was frozen to 28.51\,d \citep{andry11}. Our best fit is plotted over the data in Fig. \ref{fig:orbit5}. The results are given in Table \ref{tab:orbit5} and are discussed in section 4.

\begin{table}
  \caption{The orbital parameters for SXP8.80 from the analysis of 3--10 keV \textit{RXTE} PCA data.}
  \label{tab:orbit5}
  \centering
\begin{tabular}{|l|c|l|}
  \hline
  Parameter &  & Orbital Solution \\
  \hline
  \hline
  Orbital period & $\textit{P}_{orbital}$ (d) & 28.51 \textit{(frozen)}$^{1}$ \\
  Projected semimajor axis & $\textit{a}_{x}$sin{\it i} (light-s) & $112\pm5$ \\
  Longitude of periastron & $\omega$ ($^{o}$) & $178\pm4$ \\
  Eccentricity & $\textit{e}$ & $0.41\pm0.04$ \\
  Orbital epoch & $\tau_{periastron}$ (MJD) & $55106.7\pm0.2$ \\
  Spin period & $\textit{P}$ (s) & $8.9038\pm0.0001$ \\
  First derivative of $\textit{P}$ & $\dot{P}$ ($10^{-10}\mathrm{ss}^{-1}$) & $-6.9\pm0.6$ \\
  Goodness of fit & $\chi^{2}_{\nu}$ & 2.60 \\
  \hline
\end{tabular}
\begin{flushleft}
   $^{1}$ Orbital period frozen at the period found by \citet{andry11}
\end{flushleft}
\end{table}

\subsection{SXP74.7 = RX J0049.1$-$7250 = AX J0049$-$729}

RX J0049.1$-$7250 was discovered during a \textit{ROSAT} PSPC observation of the SMC in October 1991 \citep{kahabka96}. It appeared variable by more than a factor of 10 and was highly absorbed. \citet{kahabka96} concluded that the source was an XRB on the far side of the SMC, but could not rule out a time variable AGN. A new pulsar was discovered by \textit{RXTE} during pointed observations of SMC X-3, with a periodicity of 74.8 $\pm$ 0.4\,s \citep{cor98}. This was confirmed by the detection of a 74.675 $\pm$ 0.006\,s period during an \textit{ASCA} observation on 1997 November 13 \citep{yok98}. \citet{kahabka98} associated the \textit{ASCA} source with the highly variable \textit{ROSAT} source \citep{kahabka96} and suggested this as a Be type transient. Optical follow up work by \citet{steve99} deduced one probable and one possible counterpart to the X-ray source (which they refer to as objects 1 \& 2). Object 1 has since been confirmed as the correct counterpart from the detection of a 33.4 $\pm$ 0.4\,d period in the optical light curve \citep{schmit05}, assumed to be the orbital period of the system. \citet{mcbride08} classify the counterpart as a B3\,V emission line star.

\begin{figure}
 \includegraphics[width=64mm,angle=90]{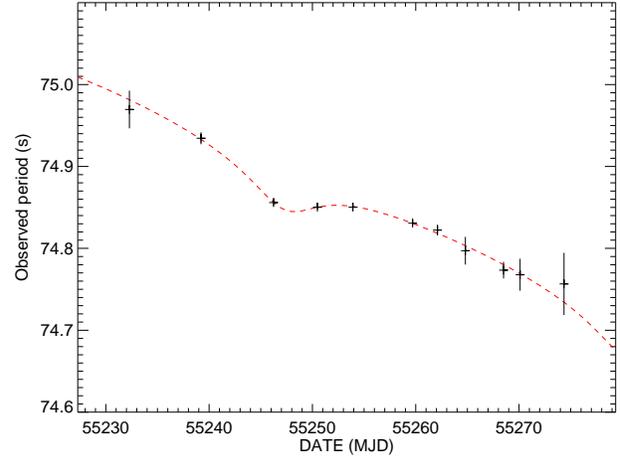}
  \caption{The spin period of SXP74.7 as measured by \textit{RXTE}. Very little orbital modulation is apparent, though the outburst is sampled well. To aid fitting, the orbital period was frozen to 33.38\,d. The fit is overplotted and is shown in Table \ref{tab:orbit6}.\label{fig:orbit6}}
\end{figure}

There have been sporadic detections of SXP74.7 with \textit{RXTE} since its discovery, although no major outburst was seen until MJD 55232. This Type II outburst was observed regularly by \textit{RXTE}, to try and follow the period evolution. Unfortunately, the outburst lasted for less than one orbit, making fitting difficult. The period history is plotted in Fig. \ref{fig:orbit6}, with the data cut at the 99.9999$\%$ (5\,$\sigma$) significance level. Besides a small kink at around MJD 55247, there is very little binary motion visible making fitting more challenging. This could be caused by a low orbital inclination with respect to our line-of-sight or the modulation could be getting swamped by an exceptionally large spin-up component. The orbital period was frozen to the value found by \citet{andry11} to help fit the data. Our best fit is plotted over the data and the parameters are given in Table \ref{tab:orbit6}. From the reduced chi-square value it is clear that the model is over-fitting the data. This is reflected in the larger errors on many of the parameters. Again, the results are discussed in the next section.

\section{Discussion}

The 4 systems presented above bring the total number of BeXRBs with reliably measured orbital parameters in the SMC to 6. As such, it is the first time in which we have a large enough sample of SMC systems to compare with the parameters of Galactic HMXBs. In this section we derive mass functions, inclination angles and orbital semimajor axes for our SMC sample and compare them and the other binary parameters to parameters calculated from studying Galactic systems.

\begin{table}
  \caption{The orbital parameters for SXP74.7 from the analysis of 3--10 keV \textit{RXTE} PCA data.}
  \label{tab:orbit6}
  \centering
\begin{tabular}{|l|c|l|}
  \hline
  Parameter &  & Orbital Solution \\
  \hline
  \hline
  Orbital period & $\textit{P}_{orbital}$ (d) & 33.38 \textit{(frozen)}$^{1}$ \\
  Projected semimajor axis & $\textit{a}_{x}$sin{\it i} (light-s) & $147\pm 15$ \\
  Longitude of periastron & $\omega$ ($^{o}$) & $186\pm18$ \\
  Eccentricity & $\textit{e}$ & $0.40\pm0.23$ \\
  Orbital epoch & $\tau_{periastron}$ (MJD) & $55280.9\pm1.5$ \\
  Spin period (MJD 55247.2) & $\textit{P}$ (s) & $74.867\pm0.002$ \\
  First derivative of $\textit{P}$ & $\dot{P}$ ($10^{-10}\mathrm{ss}^{-1}$) & $-635\pm37$ \\
  Goodness of fit & $\chi^{2}_{\nu}$ & 0.33 \\
  \hline
\end{tabular}
\begin{flushleft}
   $^{1}$ Orbital period frozen at the period found by \citet{andry11}
\end{flushleft}
\end{table}

\subsection{The Binary Mass Functions}

The masses of the two stars can be described by the X-ray and optical mass functions of the binary system:

\begin{eqnarray}
  f_{X}(M) & = & \frac{{K_{x}}^{3}P\,(1-e^{2})^{3/2}}{2\pi G} = \frac{M_{c}\sin^{3}i}{(1 + q)^{2}}
 \label{equ:xmass}
\end{eqnarray}

and

\begin{eqnarray}
  f_{C}(M) & = & \frac{{K_{c}}^{3}P\,(1-e^{2})^{3/2}}{2\pi G} = \frac{M_{x}\sin^{3}i}{(1 + 1/q)^{2}}
 \label{equ:cmass}
\end{eqnarray}

\noindent respectively. $M_{x}$ and $M_{c}$ are the masses of the NS and the optical counterpart, $i$ is the inclination of the orbital plane to the line of sight, $P$ is the orbital period, $e$ is the eccentricity and $q$ is the mass ratio (=$M_{x} / M_{c}$). The semiamplitudes of the radial velocity curves are given by

\begin{eqnarray}
  K_{n} & = & \frac{2\pi a_{n} \sin i}{P\,(1-e^{2})^{1/2}}
 \label{equ:amp}
\end{eqnarray}

\noindent where $a_{x}$ and $a_{c}$ are the semimajor axes of the ellipse travelled by the NS and the optical counterpart about the centre of mass of the system. Thus, if the radial velocity curves for both stars are known, along with the inclination angle, one can calculate the masses of the two stars to high precision. However, due to the non-eclipsing nature of these SMC binary systems, we are unable to make precise measurements of the NS masses. Instead, we can calculate the X-ray mass function for each of our systems using our orbital solutions and make estimates of the inclination and orbital size ($\simeq a_{x}$) using masses estimated from the spectral classification of the counterpart and the standard mass of a NS of 1.4 M$_{\odot}$. These results are shown in Table \ref{tab:summary} for the 6 SMC BeXRBs with orbital solutions. The range in the value of each inclination is based mostly on an uncertainty in the spectral classification of one spectral type either side of the published value. The uncertainty in the mass function also contributes to the range in inclination, albeit less significantly. The semimajor axes are estimated based on the most probable mass and radius of the primary star.

\begin{table}
  \caption{The derived X-ray mass function, orbital inclination and semimajor axis for the 4 systems studied in this paper and the 2 previously studied SMC BeXRBs. $M_{c}$ is estimated from the spectral classification of the authors referenced here using the table of luminosity and spectral type from \citet{dej87}, along with the standard mass-luminosity relationship for main sequence stars (L $\propto$ M$^{3.6}$). $f_{X}(M)$ is calculated using results from our fits and is shown in solar masses. The inclination and semimajor axis are estimated using equation \ref{equ:xmass}, $M_{c}$ and $M_{x} = 1.4 M_{\odot}$.}
  \label{tab:summary}
  \centering
\begin{tabular}{|l|l|l|l|l|l|}
  \hline
  Source & ($M_{c} / M_{\odot}$) & ($f_{X}(M) / M_{\odot}$) & $i$ $(^{\circ})$ & $a_{x}$ (R$_{\star}$) \\
  \hline
  \hline
  SXP2.37$^{1}$ & 18--26 & 1.27 $\pm$ 0.05 & 24 $^{+2} _{-3}$ & 9 \\
  SXP6.85$^{1}$ & 13--26 & 7.71 $\pm$ 0.92 & 50 $^{+20} _{-9}$ & 12 \\
  SXP8.80$^{1}$ & 13--23 & 1.86 $\pm$ 0.25 & 30 $^{+6} _{-4}$ & 13 \\
  SXP74.7$^{1}$ & 6--9 & 3.06 $\pm$ 0.94 & 55 $^{+20} _{-13}$ & 16 \\
  \hline
  SXP11.5$^{2}$ & 13--26 & 3.79 $\pm$ 0.48 & 37 $^{+11} _{-6}$ & 16 \\
  SXP18.3$^{3}$ & 8--13 & 1.43 $\pm$ 0.17 & 36 $^{+5} _{-7}$ & 10 \\
  \hline
\end{tabular}
 \begin{flushleft}
   \textit{Note}. References for the spectral classification are given: $^{1}$\citet{mcbride08}, $^{2}$\citet{town11}, $^{3}$\citet{schurch09}. The spectral type of SXP18.3 in \citet{schurch09} is estimated based on optical an NIR photometry as no spectrum allowing for classification is available.
\end{flushleft}
\end{table}

The orbital inclinations seen in this sample are as expected given the method of detection. Very low inclinations would mean the delays in pulse arrival times would not be detected, whilst very high inclinations would mean the X-ray source gets eclipsed by the primary star. More specifically, we can compare the inclinations to H$\alpha$ profiles of the Be star to investigate the (mis-)alignment of the orbital plane and the circumstellar disk. We find that SXP6.85 and SXP74.7 have quite narrow, single peaked H$\alpha$ emission, despite having the highest estimated inclinations. Conversly, we note that SXP18.3 has a lower inclination but shows prominent double peaked H$\alpha$ emission in its optical spectrum. However, we note that the mass of SXP18.3 has been estimated photometrically, not spectrally. Although this may provide some qualitative evidence of orbit-disk misalignment in these systems, more quantitative evidence may only be obtained from detailed polarimetric studies of the disk itself. The estimated semimajor axes in Table \ref{tab:summary} are quite similar and range from 9 to 16 stellar radii, but are these values what one might expect? The use of complex models that describe the dynamics and thermal structure of the circumstellar disk around the primary star helps explain what is observed here. \citet{oka07} shows that the density of the disk is several orders of magnitude lower beyond 10\,R$_{\star}$ than it is closer to the star, whilst \citet{car10} shows most of the optical and NIR flux and polarisation is emitted from within 10\,R$_{\star}$ and nearly all FIR and H$\alpha$ flux is produced inside 20\,R$_{\star}$. Thus, these models are predicting that a very large fraction of the matter in the disk is within approximately 10\,R$_{\star}$. This prediction goes some way to explaining the nature of the X-ray outbursts that we see in the SMC binary systems. SXP74.7 and SXP11.5 have the largest predicted orbital size (relative to the radius of the counterpart) and are observationally the least active of the sample, rarely undergoing a Type I outburst and only once being seen in a Type II outburst. SXP6.85, SXP8.80 and SXP18.3 have smaller predicted orbital sizes and are much more active, often being detected in Type I or Type II outbursts. This is likely due to the NS passing closer to, or further into, the circumstellar material than those in larger orbits. The exception to this seems to be SXP2.37 which has the smallest predicted orbital size, but is very rarely detected in X-ray outburst. This can be explained when considering the low eccentricity of the orbit. \citet{oka01} predict that for low eccentricity HMXBs, the circumstellar disk will get truncated at the 3:1 resonance radius, in contrast to intermediate and high eccentricity systems in which truncation is much less efficient and occurs at larger distances near the Roche lobe radius of the star. Thus, the disk in SXP2.37 could be truncated at a much smaller radius than those of the other systems in our sample, explaining the small number of X-ray outbursts seen.

\subsection{Comparision with the Galactic population}

\begin{table*}
  \caption{All high-mass X-ray binaries with reliably known orbital periods and eccentricities.}
  \label{tab:cat}
  \centering
\begin{tabular}{|l|l|l|l|l|l|l|l|}
  \hline
  Source name & Spectral type & $\textit{P}_{spin}$ (s) & $\textit{P}_{orbital}$ (d) & $\textit{e}$ & $\textit{a}_{x}$sin{\it i} (light-s) & $\omega$ ($^{o}$) & $f_{X}(M)$ (M$_{\odot}$) \\
  \hline
  \hline
  \medskip
  Be/X-ray binaries & & & & & & & \\

  PSR\,J0045$-$7319$^{1}$ & B1\,V & 0.93 & 51.169 & 0.808 & $174.235\pm0.002$ & $115.236\pm0.002$ & 2.169 \\
  RX\,J0049.1$-$7250 (SXP74.7) & B3\,V & 74.9 & $33.38\pm0.01$ & $0.40\pm0.23$ & $147\pm15$ & $186\pm18$ & $3.06\pm0.94$ \\
  2E\,0050.1$-$7247 (SXP8.80) & O9.5-B0\,IV-V & 8.89 & $28.51\pm0.01$ & $0.41\pm0.04$ & $112\pm5$ & $178\pm4$ & $1.86\pm0.25$ \\
  MX\,0053+604 ($\gamma$\,Cas)$^{2}$ & B0.5\,IV & & $203.59\pm0.29$ & $0.260\pm0.035$ & & $47.9\pm8.0$ & \\
  XTE\,J0055$-$727 (SXP18.3)$^{3}$ & B1-B3\,V & 18.4 & $17.79\pm0.01$ & $0.43\pm0.03$ & $75\pm3$ & $15\pm6$ & $1.43\pm0.17$ \\
  SMC\,X$-$2 (SXP2.37) & O9.5\,III-V & 2.37 & $18.38\pm0.02$ & $0.07\pm0.02$ & $73.7\pm0.9$ & $226\pm8$ & $1.27\pm0.05$ \\
  XTE\,J0103$-$728 (SXP6.85) & O9.5-B0\,IV-V & 6.85 & $21.9\pm0.1$ & $0.26\pm0.03$ & $151\pm6$ & $125\pm6$ & $7.71\pm0.92$ \\
  IGR\,J01054$-$7253 (SXP11.5)$^{4}$ & O9.5-B0\,IV-V & 11.5 & $36.3\pm0.4$ & $0.28\pm0.03$ & $167\pm7$ & $224\pm10$ & $3.79\pm0.48$ \\
  4U 0115+63$^{5}$ & B0.2\,V & 3.61 & 24.317 & $0.342\pm0.004$ & $140.69\pm0.72$ & $48.5\pm0.9$ & $5.06\pm0.08$ \\
  LS\,I\,+61\,303$^{6}$ & B0\,V & & $26.496\pm0.003$ & $0.537\pm0.034$ & & $40.5\pm5.7$ & \\
  V0332+53 (BQ\,Cam)$^{7}$ & O8.5\,V & 4.38 & $36.50\pm0.29$ & $0.417\pm0.007$ & $82.49\pm0.94$ & $283.49\pm0.91$ & $0.45\pm0.02$ \\
  4U\,0352+32 (X\,Per)$^{8}$ & O9.5\,III-B0\,V & 837.7 & $250.3\pm0.6$ & $0.111\pm0.018$ & $454\pm4$ & $288\pm9$ & $1.60\pm0.04$ \\
  CI\,Cam$^{9}$ & B4\,III-V & & $19.41\pm0.02$ & $0.62\pm0.01$ & $140\pm4$ & & $7.82\pm0.67$ \\
  A\,0535+26$^{10}$ & B0\,III-V & 103.5 & $110.3\pm0.3$ & $0.47\pm0.02$ & $267\pm13$ & $130\pm5$ & $1.68\pm0.25$ \\
  A\,0538$-$66$^{11}$ & B2\,III & 0.07 & 16.646 & $0.82\pm0.04$ & & $222\pm21$ & \\
  SAX\,J0635.2+0533$^{12}$ & B1\,III-B2\,V & 0.03 & $11.2\pm0.5$ & $0.29\pm0.09$ & $83\pm11$ & $356\pm24$ & $4.9\pm2.0$ \\
  GS\,0834$-$430$^{13}$ & B0-2\,III-V & 12.3 & $105.8\pm0.4$ & $0.14\pm0.04$ & $128\pm43$ & $140\pm44$ & $0.2\pm0.2$ \\
  PSR\,1259$-$63$^{14}$ & B2 & 0.05 & 1236.724 & 0.87 & & 138.665 & \\
  GRO\,J1008$-$57$^{15}$ & B1-B2\,V & 93.5 & $247.8\pm0.4$ & $0.68\pm0.02$ & $530\pm60$ & & $2.6\pm0.9$ \\
  1A\,1118$-$616$^{16}$ & O9.5\,IV-V & 407.7 & $24.0\pm0.4$ & $<$ 0.16 & $54.9\pm1.4$ & $310\pm30$ & $0.31\pm0.03$ \\
  2S\,1417$-$624$^{17}$ & B1\,V & 17.5 & $42.12\pm0.03$ & $0.417\pm0.003$ & $207.1\pm1.0$ & $298.85\pm0.68$ & $5.37\pm0.08$ \\
  XTE\,J1543$-$568$^{18}$ & Be & 27.1 & $75.56\pm0.25$ & $<$ 0.03 & $353\pm8$ & & $8.27\pm0.56$ \\
  2S\,1553$-$542$^{19}$ & Be & 9.27 & $30.6\pm2.2$ & $<$ 0.09 & $164\pm22$ & & $5.1\pm2.2$ \\
  SWIFT\,J1626.6$-$5156$^{20}$ & Be & 15.4 & $132.89\pm0.03$ & $0.08\pm0.01$ & $401\pm5$ & $340\pm9$ & $3.92\pm0.15$ \\
  GRO\,J1750$-$27$^{21}$ & & 4.45 & $29.806\pm0.001$ & $0.360\pm0.002$ & $101.8\pm0.5$ & $206.3\pm0.3$ & $1.27\pm0.02$ \\
  2S\,1845$-$024$^{22}$ & Be & 94.3 & $242.18\pm0.01$ & $0.879\pm0.005$ & $689\pm38$ & $252.2\pm9.4$ & $6.0\pm1.0$ \\
  4U\,1901+03$^{23}$ & OB & 2.76 & 22.583 & 0.036 & $106.989\pm0.015$ & $268.812\pm0.003$ & $2.578\pm0.001$ \\
  GRO\,J1944+26$^{24}$ & B0-1\,IV-V & 15.8 & $169.2\pm0.9$ & $0.33\pm0.05$ & $640\pm120$ & $269\pm23$ & $9.8\pm5.5$ \\
  GRO\,J1948+32$^{25}$ & B0\,V & 18.7 & $40.415\pm0.010$ & $0.033\pm0.013$ & $137\pm3$ & $33\pm3$ & $1.69\pm0.11$ \\
  EXO\,2030+375$^{26}$ & B0\,V & 42 & 46.021 & $0.412\pm0.001$ & $244\pm2$ & $211.3\pm0.3$ & $7.36\pm0.18$ \\
  SAX\,J2103.5+4545$^{27}$ & B0\,V & 358.6 & 12.665 & $0.406\pm0.004$ & $74.07\pm0.86$ & $244.3\pm6.0$ & $2.72\pm0.09$ \\
  \hline
  \medskip
  Giant \& Supergiant X-ray binaries & & & & & & & \\

  SMC\,X$-$1$^{28}$ & B0\,Ib & 0.72 & 3.892 & 0.0002 & 53.577 & $317\pm9$ & 10.897 \\
  2S\,0114+650$^{29}$ & B1\,Ia & & 11.598 & $0.18\pm0.05$ & & $51\pm17$ & \\
  LMC\,X$-$4$^{30}$ & O8\,III & 13.5 & 1.408 & $0.006\pm0.002$ & $26.343\pm0.002$ & & $9.893\pm0.002$ \\
  Vela\,X$-$1$^{31}$ & B0.5\,Ia & 283.5 & 8.964 & $0.090\pm0.001$ & $113.89\pm0.13$ & $152.59\pm0.92$ & $19.73\pm0.07$ \\
  Cen\,X$-$3$^{32}$ & O6-8\,III & 4.8 & 2.087 & $<$ 0.0001 & 39.661 & & $15.373\pm0.001$\\
  1E\,1145.1$-$6141$^{33}$ & B2\,Ia & 297 & $14.365\pm0.002$ & $0.20\pm0.03$ & $99.4\pm1.8$ & $308\pm8$ & $5.11\pm0.28$ \\
  GX\,301$-$2$^{34}$ & B1\,Ia & 685 & $41.498\pm0.002$ & $0.462\pm0.014$ & $368.3\pm3.7$ & $310.4\pm1.4$ & $31.14\pm0.94$ \\
  4U\,1538$-$52$^{35}$ & B0\,Iab & 526.8 & 3.723 & $0.174\pm0.015$ & $56.6\pm0.7$ & $64\pm9$ & $14.04\pm0.52$ \\
  IGR\,J16393$-$4643$^{36}$ & & 910.4 & 3.688 & $0.15\pm0.05$ & $55\pm2$ & & $13.1\pm1.5$ \\
  IGR\,J16493$-$4348$^{37}$ & B0.5\,Ib & & $6.782\pm0.002$ & $<$ 0.15 & $<$ 128 & & $<$ 49 \\
  OAO\,1657$-$415$^{38}$ & B3\,Ia-ab & 38.2 & $10.444\pm0.004$ & $0.104\pm0.005$ & $106.0\pm0.5$ & $93\pm5$ & $11.72\pm0.17$ \\
  EXO\,1722$-$363$^{39}$ & B0-1\,Ia & 414.8 & 9.740 & $<$ 0.19 & $101\pm3$ & & $11.7\pm1.0$ \\
  PSR\,J1740$-$3052$^{40}$ & B & 0.57 & 231.030 & 0.579 & 756.909 & 178.646 & 8.721 \\
  IGR\,J18027$-$2016$^{41}$ & B1\,Ib & 139.6 & 4.570 & $<$ 0.2 & $68\pm1$ & & $16.16\pm0.71$ \\
  LS\,5039$^{42}$ & ON6.5\,V(f) & & 3.906 & $0.337\pm0.036$ & & $236.0\pm5.8$ & \\
  XTE\,J1855$-$026$^{43}$ & B0\,Iaep & 361.0 & 6.072 & $0.04\pm0.02$ & $80.5\pm1.4$ & $226\pm15$ & $15.19\pm0.79$ \\
  4U\,1907+09$^{44}$ & O8.5\,Iab & 437.5 & 8.375 & $0.28\pm0.04$ & $83\pm2$ & $330\pm7$ & $8.75\pm0.63$ \\
  \hline
\end{tabular}
\begin{flushleft}
   \textit{Note}. We include all HMXBs in the Galaxy and Magellanic Clouds that we have been able to find reference to in the literature for which the orbital period and eccentricity are reliably known. If the parameter is known to better than 3 decimal places, it is truncated and the error is smaller than 0.001. References for quantities presented in the Table are given here for each individual system: $^{1}$\citet{kaspi94}; $^{2}$\citet{harm00}; $^{3}$\citet{schurch09}; $^{4}$\citet{town11}; $^{5}$\citet{raich10b}; $^{6}$\citet{arag09}; $^{7}$\citet{raich10b}; $^{8}$\citet{del00}; $^{9}$\citet{bars05}; $^{10}$\citet{finger94}; $^{11}$\citet{hut85}; $^{12}$\citet{kaar00}; $^{13}$\citet{wilson97}; $^{14}$\citet{wang04}; $^{15}$\citet{coe07}; $^{16}$\citet{staubert11}; $^{17}$\citet{raich10b}; $^{18}$\citet{icm01}; $^{19}$\citet{kelly83}; $^{20}$\citet{bay10}; $^{21}$\citet{scott97}, \citet{shaw09}; $^{22}$\citet{finger99}; $^{23}$\citet{gwm05}; $^{24}$\citet{wilson03}; $^{25}$\citet{gml04}; $^{26}$\citet{wilson08}; $^{27}$\citet{bay07}; $^{28}$\citet{raich10a}; $^{29}$\citet{grund07}; $^{30}$\citet{lev91}, \citet{vdm07}; $^{31}$\citet{krey08}; $^{32}$\citet{raich10a}; $^{33}$\citet{ray02}; $^{34}$\citet{koh97}; $^{35}$\citet{clark00}; $^{36}$\citet{thomp06}; $^{37}$\citet{cus10}; $^{38}$\citet{chak93}; $^{39}$\citet{thomp07}; $^{40}$\citet{stairs01}; $^{41}$\citet{hill05}; $^{42}$\citet{arag09}; $^{43}$\citet{cor02}; $^{44}$\citet{bay06}.
\end{flushleft}
\end{table*}

\begin{figure}
 \includegraphics[width=65mm,angle=90]{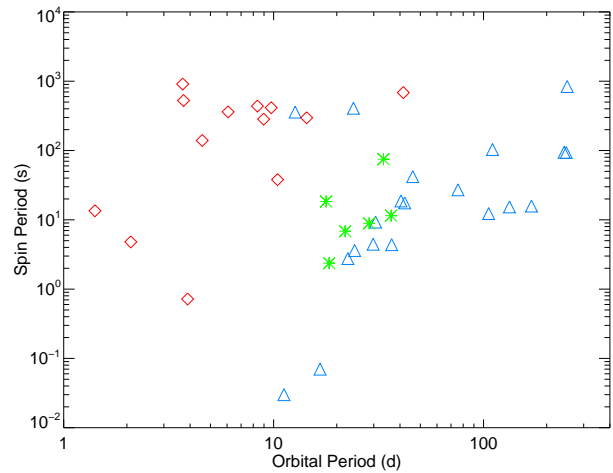}
  \caption{Corbet Diagram for all the HMXBs in Table \ref{tab:cat} that have a known spin period (excluding the PSR systems). The red diamonds represent the SG systems, the blue triangles are the Galactic BeXRBs and the green stars are the SMC BeXRBs.\label{fig:corbetdiagram}}
\end{figure}

In order to properly compare this new sample to those found in the Galaxy, a thorough review of the literature was made and a database compiled of all known HMXB systems for which the orbital period and the eccentricity are known. This list is intended to be exhaustive, although this cannot be guaranteed. In total we find 25 HMXB systems with a Be companion and 17 with a giant or supergiant companion in addition to the 6 systems discussed so far in this paper. Table \ref{tab:cat} presents all 48 HMXBs with their orbital period and eccentricity, as well as spectral type, spin period, projected semimajor axis, longitude of periastron and mass function where known. Our source list is based mostly, but not solely, on tables presented in \citet{bild97}, \citet{pfahl2002} and \citet{martin09}. We also used the updated database of BeXRBs of \citet{rag05} and the online database of \textit{INTEGRAL} sources maintained by Jerome Rodriguez and Arash Bodaghee\footnote{http://irfu.cea.fr/Sap/IGR-Sources/}. In Fig. \ref{fig:corbetdiagram} we plot the orbital and spin periods of these HMXBs (less 9 systems with unknown spin periods or PSR designations). The red diamonds represent the SG systems, the blue triangles are the Galactic BeXRBs and the green stars are the SMC BeXRBs. As expected, the SG and Be systems occupy separate parts of the plot \citep{cor86} and the SMC BeXRBs sit amongst the Galactic Be systems. This sample will be used to compare with the results from our analysis throughout the discussion.

An observation that emerges from this work, is that there are often many different periodicities found in the optical and X-ray light curves of individual sources and that these periods do not always agree. As such, great care must be taken when deciding which periodicity is the orbital period and which can be attributed to stellar pulsations, disk oscillations, superorbital periods or even aliasing between different periodicities. Fig. \ref{fig:corbetdiagram} supports our interpretation of the 21.9\,d orbital period in SXP6.85 as a longer 114\,d period would see it sit towards the edge of the BeXRB region. The 112\,d period reported by \citet{gal08} came from timing analysis of the long term \textit{RXTE} light curve. A periodogram of the updated light curve (including over 4 years of new data) refines this period to 117 $\pm$ 1\,d. The X-ray period seems to be driven by the time between outbursts, although considered with the 114\,d optically derived period and the $\sim$\,620\,d disk relaxation timescale \citep{kem08}, a superorbital period cannot be ruled out. An additional argument for the 21.9\,d period being orbital in nature is the relationship between the orbital and superorbital periods found by \citet{andry11} (see their Fig. 44). The strong correlation between these parameters suggests that for a superorbital period of $\sim$\,620\,d, the expected orbital period is around 15--20\,d. During their study, those authors conclude that the orbital period of SXP6.85 is 110\,d, making the system the largest outlier on their plot. Our new orbital period removes this outlier, further improving their correlation. SXP8.80 also shows evidence for many different periodicities in it's optical and X-ray light curves. The possible period of 185\,d reported in \citet{coe05} was only found in the MACHO red data. Those authors comment that they could not find the same period in the early OGLE data implying that the result should be taken with some caution. It is also very close to half a year -- a period that always needs to be treated with care in such data sets. The period of 33\,d reported by \citet{schmit06} is described as `weak' by those authors and possibly only seen in some early OGLE data. They accept that this could actually be the same as the 28\,d reported by \citet{cor04}. Although these periodicities could be real, possibly a signature of stellar pulsations in the Be star, it is likely that they are false. Indeed, as the optical data sets have grown in size over the years, so has the confidence in what is found. Hence, the optical period of 28\,d found by \citet{andry11} is much more likely to be real as it was found in a much longer data set. In this case, it is very likely to be the orbital period of SXP8.80 as it also agrees with the X-ray derived period of \citet{cor04}.

As mentioned previously, the low metallicity environment in the SMC may have a significant impact on the evolution of HMXBs which may manifest itself as observable differences in the population from that of the Milky Way. Fig. \ref{fig:corbetdiagram} suggests that there is little difference in the relationship between spin and orbital periods of SMC and Galactic systems. \citet{mcbride08} also show there to be no difference in the distribution of spectral types. In the following section we investigate if there is evidence of such a difference in the distribution of eccentricities and, by inferrence, supernova kick during the formation of the NS.

\subsubsection{The relationship between orbital period and eccentricity}

\begin{figure}
 \includegraphics[width=65mm,angle=90]{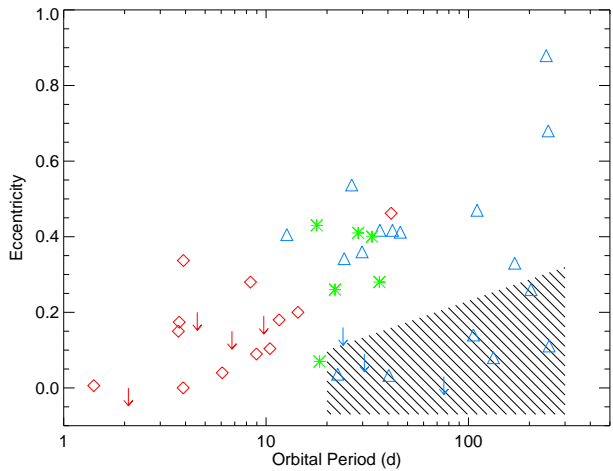}
  \caption{Orbital period against eccentricity for all the systems presented in Table \ref{tab:cat}, excluding the millisecond pulsars and CI Cam. The red diamond, blue triangle and green star symbols represent SG, Galactic Be and SMC Be systems respectively, as in Fig. \ref{fig:corbetdiagram}. The red and blue arrows are upper limits for the eccentricity of SG and Galactic Be systems respectively. The 1\,$\sigma$ errors are not plotted here for clarity, but are approximately the size of the data point or smaller. The 8 systems within the shaded region belong to the new class of low eccentricity, long orbit HMXB \citep{pfahl2002}.\label{fig:porbvecc}}
\end{figure}

The orbital solutions presented here mean that for the first time, we can investigate the relationship between the orbital period and eccentricity for a sample of SMC BeXRBs and compare this to what is seen in the Galaxy. In Fig. \ref{fig:porbvecc} we plot these two parameters for the 6 SMC sources and the Galactic sample. For clarity we remove the PSR designated systems, CI Cam and the millisecond binary pulsars A0538-66 and SAX J0635.2+0533 from our comparison as these are almost certainly not `normal' HMXBs. There are several things of note here. Firstly, there seems to be little difference between the parameter space occupied by SMC and Galactic systems. This may suggest that the metallicity difference between the galaxies is not a large factor in binary evolution, although many more orbits of SMC systems need to be solved before we can draw a firm conclusion. Secondly, the Be and SG systems occupy separate regions of the plot much like on the Corbet diagram. This is what we might expect considering the tight, circular orbits in SG systems and the wide, eccentric orbits in Be systems. The upper-right most red diamond (GX\,301-2) has shown evidence of behaving like a Be system despite its B1\,Ia classification \citep{koh97}, and so may be an exception to this relationship. Thirdly, there is a possible correlation between the two parameters, particularly when we exclude the small class of low eccentricity, long orbit OB transients \citep{pfahl2002} in the shaded region\footnote{\citet{pfahl2002} argue that these objects make up a new group of HMXB that received a much smaller kick from the supernova explosion that created the X-ray binary. This work was originally based on 6 systems, though now there are currently 8 members of this group. See \citet{gwm05} and \citet{bay10} for details on the other two systems.}. To try and quantify any possible relationship between the two parameters, a Spearman rank correlation coefficient was derived. This was preferred over a linear Pearson correlation as the relationship seems to be linear-log as opposed to simply linear. The rank correlation coefficient for the whole dataset was computed to be 0.412 with a p-value of 0.007 (significant at the 3\,$\sigma$ level). A second correlation test was performed excluding the 8 sources proposed to belong to the low eccentricity group of OB transients, resulting in a coefficient of 0.79 and p-value of $2.6 \times 10^{-8}$ (significant at the 6\,$\sigma$ level). The more significant correlation resulting from the removal of this small group of systems may reinforce the idea put forward by \citet{pfahl2002} that these are a separate population of Be systems that receive a smaller supernova kick compared to `classical' Be systems. It may also suggest that orbital period and eccentricity in HMXBs are somehow intrinsically related. However, fitting any function to the current data is difficult given the spread of values and the unknown contribution to the trend from factors such as mass, orbital size and tidal circularisation timescales. Whilst the correlation seems significant, there is an obvious alternative to this scenario. Separate correlation tests on the SG systems and Be systems yield lower correlation coefficients of 0.54 (p-value of 0.03) and 0.49 (p-value of 0.04) respectively. This means much of the overall correlation could be a result of the two groups occupying different parameter space. With the low eccentricity transients included in the Be group, this correlation coefficient drops to 0.08 (p-value 0.69). Again this favours the interpretation that we are seeing 3 different populations with different evolutionary histories, rather than one continuous population. Conversely, we can think of the weak correlations shown in both groups as evidence for a relationship between the parameters, which is only enhanced when the two groups are combined. This could be telling us something significant about the pre-supernova phase of the binary evolutionary path. \citet{del00} simulate binary periods and eccentricities resulting from SN kicks imparted onto the NS with varying initial orbital separations. Somewhat intuitively, the results show that as the initial separation is increased, a higher fraction of systems are found to have larger orbital periods and larger eccentricities. If we consider each binary system to have the same evolutionary path, the correlation in Fig. \ref{fig:porbvecc} may be telling us that SG systems generally have tighter pre-supernova orbits than Be systems. As with the question of whether there is a difference between SMC and galactic systems in this parameter space, any potential correlation in these data can only be confirmed by including more binary systems in the comparison, particularly more long orbital period Galactic systems and both long and short orbital period SMC systems. However, this will be difficult given the nature of the analysis method. Finally, we note the location of SXP2.37, immediately left of the shaded region in Fig. \ref{fig:porbvecc}. We postulate that this source is the latest member of this growing class of low eccentricity OB transient HMXB and the first member from outside the Galaxy.

\section{Summary}

The orbital parameters of four BeXRB systems in the SMC have been derived, taking the total number of extra-Galactic BeXRBs with orbital solutions to six. We find one new orbital period and confirm a second. Two further fits were made possible by freezing the orbital period parameter to known values from analysis of optical light curves. We find that one system (SXP2.37 = SMC X-2) is consistent with being part of the small class of HMXBs with low eccentricities and long orbital periods proposed by \citet{pfahl2002}, or at least lies between this class and the `normal' population. On comparing the SMC sample to Galactic systems we find there to be little difference in their binary parameters, suggesting that the low metallicity environment in the SMC does not contribute strongly to the evolution of the binary. The different types of HMXB are found to occupy different regions on a plot of orbital period and eccentricity, similar to what is seen in the Corbet diagram. A possible correlation between the two parameters is noted, although it is unclear whether this is an effect of the position of the groups on the plot or if it demonstrates a physical connection between the parameters in HMXB systems. The data clearly suffer from selection effects, in particular the lack of many long orbital period systems that will ultimately determine the answers to the questions posed here. As a by-product of our work, we present a catalogue of the orbital parameters for every HMXB in the Galaxy and Magellanic Clouds for which they are known.

\section*{Acknowledgements}

LJT is supported by a Mayflower scholarship from the University of Southampton. We would like to thank the anonymous referee for their helpful and constructive comments.

\bsp

\label{lastpage}

\end{document}